\documentstyle[psfig]{mn}

\def\be{\begin{equation}}
\def\ee{\end{equation}}

\def\msun{M_{\odot}}

\def\ergsec{\rm \  \ erg \ s^{-1}}

\date{}
\title{Possible evidence for the disk origin for the powering of jets in 
Sgr A$^*$ and nearby elliptical galaxies}
\author[Feng Yuan] 
{\parbox[]{6.in} {Feng Yuan$^{1,2,3}$ \\
\footnotesize
1. Department of Astronomy, Nanjing University, Nanjing 210093,
China. Email: fyuan@nju.edu.cn\\
2. Beijing Astrophysical Center, Beijing, 10080, China\\
3. Korea Institute for Advanced Study, Seoul, 130-012, Korea\\}}

\date{Accepted .
      Received ;
     in original form }

\pagerange{\pageref{firstpage}-- \pageref{lastpage}}
\pubyear{2000}

\begin{document}

\maketitle

\label{firstpage}

\begin{abstract}
Recent VLBA observation indicates the existence 
of an elongated (jet) structure in the compact radio source Sgr A$^*$.
This is hard to explain in the context of advection-dominated accretion 
flow (ADAF) model for this source. On the other hand, 
the mass accretion rate favored by ADAF
is 10-20 times smaller than that favored by the hydrodynamical 
simulation based on Bondi capture. 
If the latter were adopted, the predicted 
radio flux would significantly exceed the observation.
A similar situation exists in the case of nearby giant ellipticals, 
where the canonical ADAF model -- the widely assumed 
standard model for these sources -- also significantly overpredict
the radio flux.
Based on these facts,
in this paper we propose a truncated ADAF model for Sgr A$^*$ and 
three ellipticals M87, NGC 4649 and NGC 4636.
We assume that the accretion disk is truncated at a certain radius
$R_{\rm tr}$ within which the jet forms by extracting the energy of the disk. 
The radio flux is greatly suppressed due to the radiative 
truncation of the disk and the fits to the observational data are excellent. 
For example, for Sgr A$^*$, the model 
fits the observational spectrum very well from
radio including the ``excess'' below the break frequency 
to hard X-ray under a high accretion rate near the simulation value,
and the predicted  
size-frequency relationship is also in excellent agreement with 
the observation; for M87, the predicted upper limit of the jet location
is $24 R_{\rm g}$, in excellent agreement with the observational result 
that the jet is formed on scales smaller than $30 R_{\rm g}$, 
and the $\approx 20 \%$ variability at $\sim$ 1 Kev -- which is hard to 
explain in another model succeeded 
in explaining the low radio flux of M87 -- is also marginally interpreted. 
The success of the model supplies a possible evidence for the 
disk rather than the hole origin for the powering of jets. 
\end{abstract}

\begin{keywords}
accretion, accretion disks -- black hole physics -- 
galaxies: active  -- galaxies: individual 
(M87, NGC 4649, NGC 4636) -- galaxies: jets -- Galaxy: centre -- hydrodynamics
\end{keywords}

\section{Introduction}
The energetic radio source Sgr A$^*$ located
at the center of our Galaxy is widely believed to be a black hole
with mass $M = 2.5 \times 10^6 \msun$ (Mezger, Duschl \& Zylka 1996). 
Due to its proximity, the observational data are abundant, which makes it an 
ideal laboratory to test our accretion theories. Any successful model needs 
to explain its relative low-luminosity, 
the emergent spectrum from radio to X-ray wavelength, 
and the frequency-size relationship.

Several models have been put forward for Sgr A$^*$. They include
Bondi accretion (Melia 1992; 1994), emission from mono-energetic 
electrons (Duschl \& Lesch 1994), jet model 
(Falcke, Mannheim \& Biermann 1993; Falcke \& Biermann 1999),
and the most recent two temperature 
advection-dominated accretion flows (ADAFs; 
Narayan, Yi \& Mahadevan 1995; Manmoto, Mineshige, \& Kusunose 1997; 
Narayan et al. 1998). 
Among them, ADAF model is the most dynamically 
self-consistent. However, two puzzles still remain in this model. First, 
as pointed out by Quataert \& Narayan (1999), 
the mass accretion rate required in this model 
is more than 10 
times smaller than that favored by three-dimensional hydrodynamical 
simulation based on the Bondi capture, 
with the former being
$ \approx 6.8 \times 10^{-5} \dot{M}_{\rm Edd}$ (Quataert \& Narayan 1999) and 
the latter $ \approx 9 \times 10^{-4}
\dot{M}_{\rm Edd}$ (Coker \& Melia 1997). Here $\dot{M}_{\rm Edd}
=2.2 \times 10^{-8} {\rm M}{\rm yr}^{-1}$
denotes the Eddington accretion rate and $M$ is the mass of the centre hole. 
Very recent Chandra observations of Sgr A$^*$ show that its $0.1 - 10$ 
kev luminosity is $L_X \approx 4 \times 10^{33} \ergsec$ (Baganoff et al. 2000),
considerably smaller than the upper limit used by Quataert \& Narayan (1999).
Taking this into account, the accretion rate required in ADAF model would be
smaller hence the discrepancy between the two accretion rates 
will be larger. If a higher accretion rate were adopted in an ADAF model,
the bremsstrahlung and synchrotron radiations would yield an X-ray
and a radio fluxes well above the observational
limits.

Yuan (1999) find that the outer boundary condition 
of the accretion flow plays an important role in
the dynamics and radiation of an optically thin accretion flow.
This point was neglected in previous ADAF models in the literature.
After considering this effect, it was found 
that since the specific angular momentum of the flow at 
the outer boundary is low, $\Omega_{\rm out} \sim
0.15 \Omega_{\rm Kepler}$ as shown by the numerical simulation
(Coker \& Melia 1997), the accretion in Sgr A$^*$ 
should belong to Bondi-like (not Bondi type),
characterized by a much larger sonic radius and a low 
surface density compared with the conventional disk-type ADAF.
The bremsstrahlung radiation therefore is 
greatly suppressed (Yuan et al. 2000). However,  
the synchrotron radiation would still produce  
too much radio flux were the mass accretion rate favored by 
the simulation adopted.

The second puzzle is that ADAF model strongly 
underpredicts the radio flux in the radio spectrum below a ``break'' 
frequency (Narayan et al. 1998). 
Mahadevan (1998) argued that the charged pion produced by the energetic
proton collisions in ADAF will subsequently decay into $e^-$$e^+$,
which will further produce synchrotron emission
and could serve as the origin of the radio excess below the break.
However, the result depends on the power-law index 
of the proton energy spectrum, which we are not clear.
In fact, due to the existence of the threshold energy
needed for pion production, these extra photons should be principally 
produced in the innermost region of the disk. 
This is in conflict with the observed intrinsic size 
of $\sim 72 R_{\rm g}$ at 43 GHz (Lo et al. 1998; see also Figure 2),
here $R_{\rm g}= 2GM/c^2$ is the Schwarzschild radius of the black hole.

More importantly, recent excellent near-simultaneous VLBA measurements
show that the intrinsic source structure at 43 GHz is 
elongated along an essentially north-south direction, with an
axial ratio of less than 0.3 (Lo et al. 1998). 
This is hard to explain in an ADAF model, while 
it provides a convincing evidence for the existence of a jet
in the jet-nozzle model of Falcke et al. (1993). 
In this model the radio spectrum below the break is produced by
a jet while those above the break is assumed to be produced by a 
``nozzle'' located at the lower base of the jet. 
The question is, however, this model can not account for the entire 
spectral energy distribution from radio to X-ray, and we lack
any detailed understanding to the ``nozzle'' which is crucial to the 
fit of the spectrum above the break. Considering this situation,
recently Falcke (1999) and Donea, Falcke, \& Biermann (1999)
suggested that maybe we should combine the jet model
with an accretion model, e.g., ADAF, Bondi-type accretion or others.

A similar situation exists in the case of
nearby elliptical galaxies. The existence of a massive black hole 
and the hot gas pervading the galactic center is in sharp contrast 
to their low luminosity. Recently it was suggested that ADAF 
could serve as a feasible explanation for the quiescence of the 
galaxies (Fabian \& Rees 1995). Recent observation, however, shows that 
the canonical ADAF model overpredicts the radio flux by more than two 
orders of magnitude (Di Matteo et al. 1999). The current explanation is 
to drive a wind from the accretion disk, assuming 
the accretion rate $\dot{M}=(R/R_{\rm out})^p\dot{M}_{\rm out}$ 
(Di Matteo et al. 2000),
where $R_{\rm out}$ is the radius from which the 
wind starts to play an important role and $\dot{M}_{\rm out}$ is 
the mass accretion rate there.
Since the electron is basically adiabatic, its temperature is determined 
by the compression work. When a strong wind is introduced, 
the density profile would become flatter which results in a 
significant decrease of the temperature and further the synchrotron radiation. 
The presence of a strong wind is based on the assumption that 
the Bernoulli parameter of an ADAF is positive therefore the flow
is in some sense unstable (Blandford \& Begelman 1999).
As shown by Nakamura (1998) and Abramowicz, Lasota, \& Igumenshchev 
(2000), however, the Bernoulli parameter is negative in most 
of the reasonable cases. Numerous numerical simulations also confirm the 
absence of winds from the disks (but jets do present in the 
innermost region of the disk in some cases) (Eggum, Coroniti, \& Katz 1988; 
Molteni, Lanzafame, \& Chakrabarti 1994; Igumenshchev \& Abramowicz 1998;
Stone, Pringle, \& Begelman 1999; Koide et al. 2000). 
Thus the scenario of strong winds from ADAFs
might be questionable.
Even so, the fit is not very satisfactory in some aspects, as illustrated below.
What is interesting is, 
all the sources in Di Matteo et al. (2000)'s sample have the extended 
and often highly collimated radio structures
which are again the evidence for the existence
of jets (e.g. Stanger \& Warwick 1986).

Based on the above facts, in this paper, we propose an ADAF-jet model for
Sgr A$^*$ and nearby elliptical galaxies.

\section{Truncated ADAF by a jet}

Jets are observed in many classes of astrophysical objects
involving an accretion flow, ranging from AGNs and quasars,
to old compact stars in binaries, and to young stellar objects.
There are currently two main energetic mechanisms for jets
(see recent reviews by Lovelace, Ustyugova \& Koldota 1999 and Livio 1999).
The first one is the Blandford-Znajek mechanism (Blandford \& Znajek 1977).
The rotation of the black hole drags the inertial
frame near it and twists the magnetic field line supported
by the surrounding disk. The resulted magnetic stress is then released
as a Poynting flux away from the hole. In this mechanism,
the powering of jets is provided by the rotating hole.
The second is the 
electromagnetic output from the inner region of the 
accretion disk (Blandford 1976; Lovelace 1976;
Blandford \& Payne 1982; see Spruit 1996 for more references). 
The twisting of the field caused by
the rotation of a Keplerian accretion disk results 
in outflows out of the disk,
or the material in the disk is accelerated by the
centrifugal force and magnetic pressure gradient force. In these models,
the powering of jets is the energy output from the disk.
We are not clear at present which mechanism plays the dominated role even
after so many years' efforts.

But we are sure at this point that the jet must originates
from the inner region of the disk. This conclusion is based on
numerous observations. One evidence is that
the jet velocity is always of the order of the escape velocity from the 
central objects in any case involving jets. 
Another good example comes from the observation 
of the jet in M87, as to be stated
in this paper (for more evidence, see Livio 1999). Numerical simulations
also obtain the same result (e.g., Koide et al. 2000).

For our purpose, in this paper we adopt the ``disk'' origin of jets.
We assume the jet forms from a certain radius $R_{\rm tr}$ by 
extracting the energy of the ADAF. Since the jet power is generally comparable
with the total accretion power of the disk (Rawlings \& Saunders 1991), 
we expect that winthin $R_{\rm tr}$ 
a large fraction of the energy (and mass, maybe) 
of the disk will be transferred 
into the jet. The detailed physical mechanism
is very complicated and is an open problem. According to the numerical 
simulation, in addition to the 
electromagnetic process, hydrodynamic
adiabatic expansion may also play a role (e.g. Koide et al. 2000). As a result,
the temperature of the disk within $R_{\rm tr}$ will significantly decrease
compared with an ADAF without jet.
Since the synchrotron radiation, which is the dominant radiation process
in the innermost region of ADAF (the bremsstrahlung emission can be
neglected here), is very sensitive to the temperature,
$L_{\nu} \propto T_e^{21/5}$ for the ``general'' frequency and
$\nu L_{\nu} \propto T_e^7$ for the peak frequency (Mahadevan 1997),
we here for simplicity assume that within $R_{\rm tr}$
the radiation of the disk (if it still exists) is completely suppressed,
i.e., ADAF is radiatively truncated by the jet. 

The resulted spectrum of the truncated ADAF can be qualitatively understood 
by the following simple estimate. The synchrotron photons 
emitted at a radius $R$ in an ADAF,
which are responsible for radio spectrum, are highly self-absorbed and
give a blackbody spectrum, up to a critical frequency $\nu_c(R)$ above
which the radiation becomes optically thin.
The emergent spectrum of the ring at radius $R$ 
is peaked at this frequency and the total radio spectrum is the superposition
of each ring of ADAF. Therefore each point along the radio spectrum
corresponds to a certain radius in an ADAF (Mahadevan 1997), 
\be
\nu_c(r) \approx 30 m_9^{-1/2}\dot{m}_{-4}^{1/2}T_{e9}^2 r^{-5/4} {\rm GHz}
\ee
Here, $m_9=M/(10^9\msun)$, $\dot{m}_{-4}=\dot{M}/(10^{-4}\dot{M}_{\rm Edd})$,
$T_{e9}=T_e/10^9{\rm K}$, and $r=R/R_{\rm g}$.
Therefore, the truncation of ADAF at $R_{\rm tr}$ will
``take away'' the high-frequency part of the radio spectrum 
from $\nu(R_{\rm tr}/R_{\rm g})$ up to the previous
radio peak $\nu(r=1)$, as illustrated by the comparison between the
thin solid and dotted lines in Figure 1.
Thus the radio flux is strongly suppressed.

The procedure to calculate the spectrum of the truncated ADAF
is as follows. 
Bremsstrahlung and synchrotron radiations amplified by Comptonization
are taken into account and two-temperature plasma assumption is adopted.
For simplicity we set the outer boundary 
at $10^4 R_{\rm g}$ and
the temperatures of electrons and ions 
at $R_{\rm out}$ are set as $T_{\rm i,out} \approx
T_{\rm e,out} \approx 7 \times 10^7 {\rm K}$. 
Strictly, we should set the outer boundary at the radius 
where the accretion begins, and adopt the flow's 
temperature and angular momentum there as the outer boundary condition,
like in the case of Sgr A$^*$ in Yuan et al. (2000).
Our previous calculation to Sgr A$^*$ shows
that if only we set the outer boundary far enough from the hole, 
say $\ga 10^4 R_g$ in the present case,
the effects due to different $T_{\rm out}$ are not large.
However, the angular momentum of the accretion flows
in both Sgr A$^*$ and ellipticals is very small,
we therefore set $\Omega_{\rm out} \la 0.2\Omega_{\rm Kepler}$ in 
all the cases thus the accretion mode is 
of Bondi-like under the parameters we adopt.
Other parameters we generally set are $\alpha=0.1$ and 
$\beta=0.5$. Here $\alpha$
is the viscosity parameter in Shakura \& Sunyaev-type viscous
description and $\beta$ denotes the ratio of 
the gas pressure to the sum of the magnetic and gas pressures. 
Note since we assume a strict equipartition between the magnetic pressure
and the gas pressure, the truncation radii we obtain below 
are all upper limits.
We self-consistently solve the set of equations describing the 
radiation hydrodynamics of the accretion flow,
as we did for a ``not-truncated'' ADAF in Yuan et al. (2000), 
i.e. we require the solution must satisfy 
the outer boundary condition at $R_{\rm out}$ and the sonic point condition
at a sonic point. This is
reasonable because the flow outside of $R_{\rm tr}$ cann't sense the
formation of the jet in the {\em supersonic} 
part of the accretion flow. However, 
since the disk is radiatively truncated at $R_{\rm tr}$,
when we calculate the 
emergent spectrum, we only integrate
from the outer boundary 
to the truncation radius $R_{\rm tr}$ rather than to the horizon.
$R_{\rm tr}$ is treated as a free parameter in our model.

For completeness, for some sources we calculate the radiation of the jet
as well. However, compared with ADAF, there are 
much more free parameters in 
the general jet model developed especially for BL Lac 
objects (e.g. Ghisellini, 
Maraschi, \& Treves 1985). Here 
we follow Falcke \& Biermann (1999) to calculate the radiation of the jet.
The formula are derived from the Blandford \& Konigl (1979) model which 
calculate the synchrotron emission of a freely expanding, pressure 
driven jet as a function of jet power.
The free parameters include the jet power $Q_{\rm jet}$, 
the characteristic electron Lolentz factor
$\gamma_e$, and the angle between the sight line and the jet axis
$i$. We set the scale height
of the acceleration zone $Z_{\rm nozz}$ -- a free parameter in 
Falcke \& Biermann (1999)'s model -- to equal to the 
scale height of the disk at $R_{\rm tr}$. 
We should note that the formula we adopt is somewhat simplified.
For example, it assumes the electron energy distribution is
quasi-monoenergetic rather than the more realistic 
power law. Since in almost all practical cases, the synchrotron radiation 
is highly self-absorbed, the final spectrum results from the superposition
of the radiation from different radii. 
Therefore, even though the energy distribution were
power law, the result should be roughly the same if only the maximum 
Lolentz factor is not too large, which is reasonable if the radiative
zone is not too far away from the base of the jet hence the electrons have
not been accelerated to a power law by, e.g. a shock. 
We sum the radiation of the disk and the jet to obtain the total spectrum.

\section{Applications to individual sources}
\subsection{Sgr A$^*$}

\begin{figure}
\psfig{file=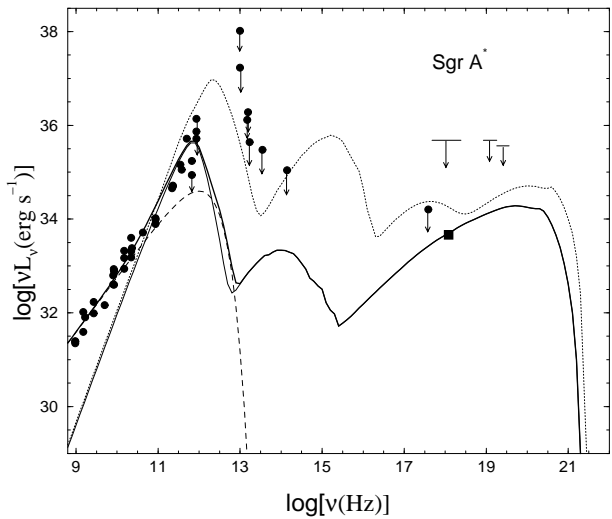,width=0.45\textwidth,angle=270}
\caption{The predicted spectrum of the ADAF-jet model for 
Sgr A$^*$ together with the observation taken from 
Narayan et al. (1998) 
except the new Chandra data denoted by a filled square.
The dotted line is for the ``not-truncated''  
ADAF with the mass accretion rate 
$1.5 \times 10^{-4} \dot{M}_{\rm Edd}$.
The thin-solid line is
for the truncated ADAF with the same accretion rate, the truncation radius
is $R_{\rm tr}=10 R_{\rm g}$.
The dashed line is for the jet and the thick-solid line 
is the sum of the truncated ADAF and the jet. 
Other parameters are $M=2.5 \times 10^6 \msun,
\gamma_e=125, i=45^{\circ}$, and $Q_{\rm jet}=10^{38.4} \ergsec$.}
\end{figure}

Figure 1 shows the calculated spectrum for Sgr A$^*$ together with 
the observational data taken from Narayan et al.(1998) except
that the new Chandra data denoted by a filled square is taken from
Baganoff et al. (2000). The thin solid
line is for a truncated ADAF with $\dot{M}=1.5 \times 10^{-4} 
\dot{M}_{\rm Edd}$, the dashed line is for
a jet, and the thick-solid line denotes the total 
spectrum of the ADAF-jet system. The truncation radius 
$R_{\rm tr}= 10R_{\rm g}$. 
The dotted line is for the ``not-truncated''
ADAF with the same parameters and outer boundary conditions except 
$\beta=0.9$.
Even though we adopt such a weak magnetic field,
we find that the synchrotron process still emits too much radio flux and
the Comptonization of the synchrotron photons
produces a X-ray flux well above the observational measurement. 
Compared with the ``not-truncated''
ADAF, the radio flux is significantly suppressed in the truncated ADAF
and the X-ray emission is also reduced due to the reduction
of Comptonized synchrotron photons.
From the figure we find that the spectrum below $\sim 50$ GHz is mainly
contributed by the jet while those above is principally
produced by the disk and the fit to the observation
is good.

Although the accretion rate favored in our model is $\sim 6$ times
smaller than the numerical simulation of Coker \& Melia (1997), 
we note that such a discrepancy could be
reduced by $\sim$ 3 times if we adopted a larger viscous parameter 
$\alpha=0.3$ because of the scaling law $\rho \propto \dot{M}/\alpha$.
In this case, the viscous dissipation will produce more energy hence
the electron temperature will increase. As a result,
more radio flux will be emitted by the synchrotron process.
Then the conflict between the ``not-truncated'' ADAF and the observation
will be more serious, while in the truncated ADAF model, a slightly larger
truncation radius $R_{\rm tr}$ is then required to give a satisfactory fit.

Figure 2 shows the
predicted size-frequency relationship of our ADAF-jet model together
with the observational data taken from Lo et al. (1998) (for 43 GHz) and
Krichbaum et al. (1998) (for 86 and 215 GHz) 
\footnote{Because of the existence of two components, 
jet and disk, and the reason stated in Krichbaum et al. 1998,
their ``case 2'' result is more reasonable and is adopted here.
See Krichbaum et al (1998) for details}. 
The solid one is the result of our ADAF-jet model, with the line
below and above the break frequency at around 
$\sim 50$ GHz resulted by the jet and
the truncated ADAF respectively.
For comparison, we also show by the dashed line 
the prediction of a best-fit ``not-truncated'' ADAF model 
similar with that in Quataert \& Narayan (1999),
\footnote{In addition to the difference of the mass accretion rate,
the different
prediction of the ``not-truncated'' ADAF between
the present result (dashed line) and that in Narayan et al. (1998) is
due to the difference of the adopted electron adiabatic index $\gamma_e$.
We take it to be that of a monatomic ideal gas, as in
Quataert \& Narayan (1999), while
in Narayan et al. (1998) $\gamma_e$ includes the
contribution of the magnetic density as well. Compared with ours,
the choice in Narayan et al. (1998) results in
a lower electron temperature therefore
a larger radio source size is needed
to produce the observed flux at a given frequency.
Due to the same reason, in Quataert \& Narayan (1999) and the present paper,
a noticeably smaller mass accretion rate
than Narayan et al. (1998)
is required to fit the observation.
However, as argued by Quataert \& Narayan (1999),
$\gamma_e$ should not include the contribution of the magnetic density.}
, but with a smaller
accretion rate of $\dot{M}=5.7 \times 10^{-5} \dot{M}_{\rm Edd}$ because
of the new Chandra upper limit. Other parameters are
$\alpha=0.1, \beta=0.5, T_{\rm i, out} \approx T_{\rm e,out}
\approx 7 \times 10^7 $K, and 
$\Omega_{\rm out} \approx 0.45 \Omega_{\rm Kepler}$.
Note in this ADAF model, $\Omega_{\rm out}$ is large
so it is of disk-like hence $\dot{M}$ is smaller than our Bondi-like model.
We see that the two models are both compatible with the data at 86 and 215 GHz, 
but our ADAF-jet model gives a better fit to the observation
at 43 GHz because of the inclusion of the jet. In this context,
we note that although the ``ADAF+wind'' model of Quataert \& Narayan (1999) 
could also interpret the spectrum, 
even though the high accretion rate favored by simulation were
adopted, the predicted size-frequency relationship would be in conflict
with observation at 43 GHz, since that model should produce a
similar frequency-size relationship with the ``not-truncated'' ADAF model. 

\begin{figure}
\psfig{file=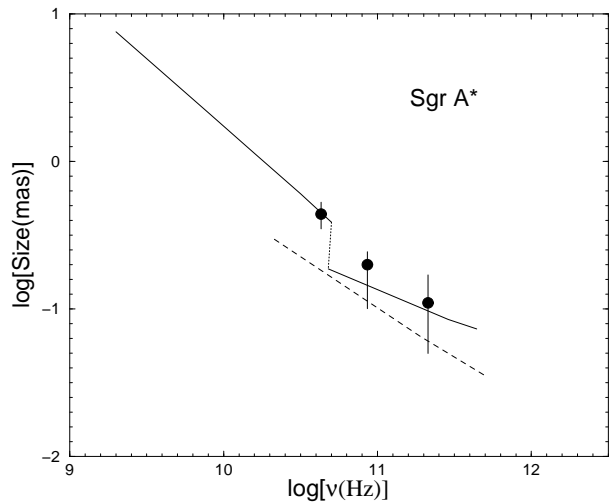,width=0.45\textwidth,angle=270}
\caption{The predicted size-frequency relationship together with 
the observation for Sgr A$^*$. 
The thick solid line is the 
result of our ADAF-jet model with the line above the vertical dotted line  
being for the jet and the line below for the truncated ADAF. The
long-dashed line is for a ``not-truncated'' ADAF. A distance of 8.5 kpc
is assumed.}
\end{figure}

\subsection{Nearby Elliptical Galaxies}

Encouraged by the above success, 
we further apply our model to the six ellipticals presented 
in Di Matteo et al. (2000), where jets also exist.
Due to the successful applications
of Falcke's jet model to other sources like M81, GRS 1915+105,
and NGC 4258 (Falcke \& Biermann 1999), 
we assume here that for our purpose 
this model can give an enough 
description to the weak jets (excluding M87) in ellipticals as well.
We find that due to the suppression of the radio flux 
by the truncation of the disk, 
the predicted spectra are in good agreement with the observations.
We present below three sources as illustrations. The observational 
data are taken from Reynold et al. (1996) (for M87) and 
Di Matteo et al.(1999) (for NGC 4649 and NGC 4636). 

\subsubsection{NGC 4649}

\begin{figure}
\psfig{file=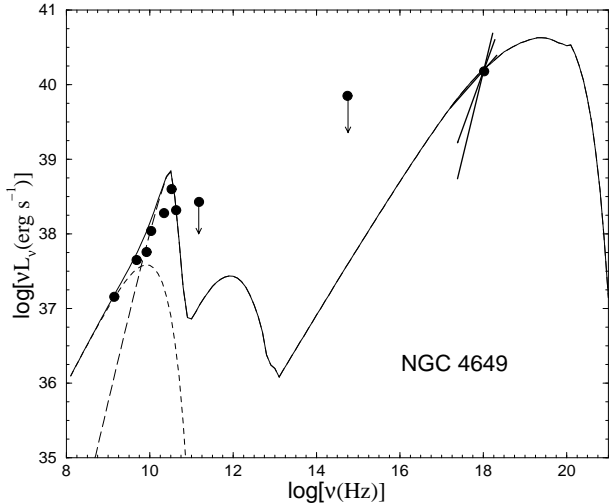,width=0.45\textwidth,angle=270}
\caption{The predicted spectrum of the ADAF-jet 
model for  NGC 4649 together with the observation. The long-dashed 
line is the result of the truncated ADAF, the short-dashed 
line is for the jet, and the solid line shows the sum. 
The parameters are $M=4 \times 10^9 \msun, \dot{M}
=8 \times 10^{-3} \dot{M}_{\rm Edd}, Q_{\rm jet}=
10^{40.7} {\ergsec}, i=20^{\circ},
R_{tr}=40 R_g$, and $\gamma_e=200$.}
\end{figure}

We first model NGC 4649 because it has the best 
observational constraints in 
Di Matteo et al. (2000)'s sample. 
Figure 3 shows the predicted spectrum of this source 
together with the observation. The long-dashed 
line is the result of the truncated ADAF disk, the short-dashed 
line is for the jet, and the solid line shows the sum. 
As clearly shown, this source, which can be matched  
well by the ADAF with winds, also can be matched quite well by a truncated ADAF.
However, in the wind model,
if we assume the wind were to start at the outer 
boundary $10^4R_{\rm g}$, the radius
where the accretion starts, a fairly strong suppression
of the X-ray emission above $\sim$ 5keV
is introduced, which is in conflict with the ASCA data (Di Matteo et al. 2000).
As stated by Di Matteo et al. (2000), this is because 
the introduction of winds at $10^4R_{\rm g}$
strongly decreases the bremsstrahlung emissivity at $10^3 \sim 10^4R_{\rm g}$,
where the X-ray emission $\ga 10$ keV mainly originates from.
Therefore, in their model, they have to assume the winds to start
at a certain radius much smaller than $10^4R_{\rm g}$, which
is not easy to understand. \footnote{Although they argue that if the 
specific angular momentum of the flows at the accretion radius
$\sim 10^4R_{\rm g}$ is a small fraction($\xi$) of the corresponding Keplerian
value, then the angular momentum will start to dominate the flow at 
$r \sim \xi^2 10^4R_{\rm g}$, where the winds start to play an
important role. However, according to our numerical calculation 
(Yuan et al. 2000), when the specific angular momentum at the 
accretion radius is rather low, it will remain highly sub-Keplerian
throughout the way to the hole rather than dominate the flow at a certain small
radius. This is the so-called Bondi-like accretion. 
See also Abramowicz (1998).}
In our model, the X-ray emission above $\sim 5$keV doesn't show any suppression,
in agreement with ASCA observation. This is because out of the (small)
truncation radius  $R_{\rm tr}$, ADAF is the canonical one in the sense 
that no winds are introduced.
In addition, due to the introduction 
of the jet, the radio flux near 1 GHz, which 
ADAF with winds model underpredict it greatly, can be interpreted
as the contribution from the jet, as in the case of Sgr A$^*$.

\subsubsection{NGC 4486 (M87)}

\begin{figure}
\psfig{file=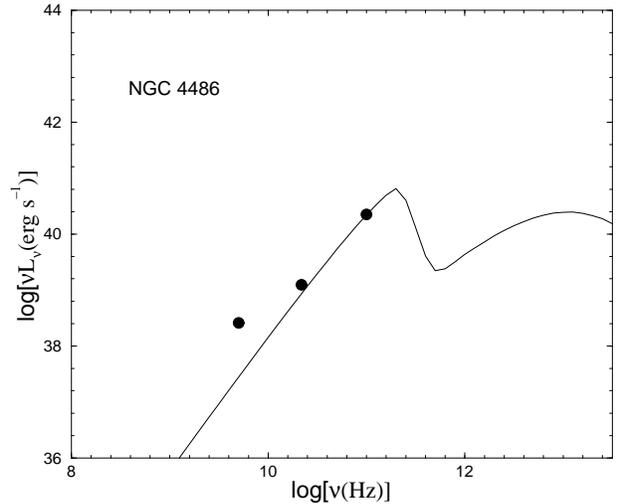,width=0.45\textwidth,angle=270}
\caption{The predicted spectrum of the truncated ADAF for 
NGC 4486 (M87) together with the observation. 
The parameters are $\dot{M}
=3 \times 10^{-2} \dot{M}_{\rm Edd}$ and $M=3 \times 10^9 \msun$. 
The truncation radius $R_{\rm tr}$ equals $24 R_g$.}
\end{figure}

This is the only source in our examples for which we have 
clear observational evidence
for the existences of a strong jet and a disk. Its jet is famous, while HST 
spectroscopy of its nucleus has given
strong evidence for a rapidly rotating ionized gas disk
at its center (e.g., Ford et al. 1994). 
Although its high frequency radio data is consistent
with the canonical ``not-truncated'' ADAF model, the 
X-ray emission in this case 
is due to Comptonization of the synchrotron photons and 
the predicted slope is too soft to be consistent with the ASCA data.
Therefore the radio flux must also be significantly smaller than the 
prediction of a canonical ADAF (Di Matteo et al. 2000).
Figure 4 shows the predicted radio spectrum of 
this source by the truncated ADAF
together with the observation. 
Three points need to be emphasized. 
First, compared with the wind model, the truncated ADAF 
can  fit the 100GHz VLBI data much better.
Although the 5 GHz VLBI data is still strongly underpredicted, 
the excess may again be due to the contribution of
the jet, like in the cases of Sgr A$^*$ and NGC 4649.
Second, both winds and the truncated 
ADAF models can give a good match to the spectrum of this source (and others),
but they are intrinsically different. 
One point reflecting the difference is that, 
the wind starts at a much larger radius, 
typically several hundreds of $R_{\rm g}$,
while in the latter, the theoretically predicted truncation radius at which 
the jet forms is much smaller.
For M87, this radius equals $24 R_{\rm g}$. Were it larger, the ADAF would
predict a radio flux well below the observed one. On the other hand,
as a triumph of 
precision astronomy, the excellent 43GHz observation for this source by 
Junor, Biretta \& Livio (1999) indicates that the jet  
is formed on scales smaller
than $\sim 30 R_{\rm g}$ from the black hole, 
in excellent agreement with our prediction. 

The last point we want to emphasize concerns the variability of M87.
Another intrinsic difference between the wind model and ADAF-jet model
is that the density of the disk at moderate radius differs greatly 
under a same accretion rate.
Due to the large mass loss through the wind, the surface density
of the disk in the wind model is much smaller than the latter.
As a result, the bremsstrahlung emissivity within the wind-starting 
radius is strongly reduced. Therefore, the radius near which most of the 
bremsstrahlung emission flux at frequency $\nu$ can be produced -- $r_{\nu}$ 
-- increases and the dynamical time scale relevant to the bremsstrahlung 
variability at $r_{\nu}$ -- $t_d(r_{\nu})$ -- increases. 
Since for variability we have,
\be
\frac{\delta L_{\nu}(\delta t)}{L_{\nu}} \sim A \, {\rm Min} \left(
1,\left[ \frac{\delta t}{t_d(r_{\nu})}\right]^{3/2} \right),
\ee
thus, on timescale of $\approx$ 6 months to a year
only $\la 1\%$ variability is expected, therefore, the $\approx 20\%$ 
variability observed  by the ROSAT HRI ($\sim 1$ keV) of the core of M87
is hard to reconcile in a wind model (Di Matteo et al. 2000),
while it can be marginally interpreted if
the jet originates at $\sim$
$30 R_{\rm g}$. This is the very case of our truncated ADAF model. 

\subsubsection{NGC 4636}
\begin{figure}
\psfig{file=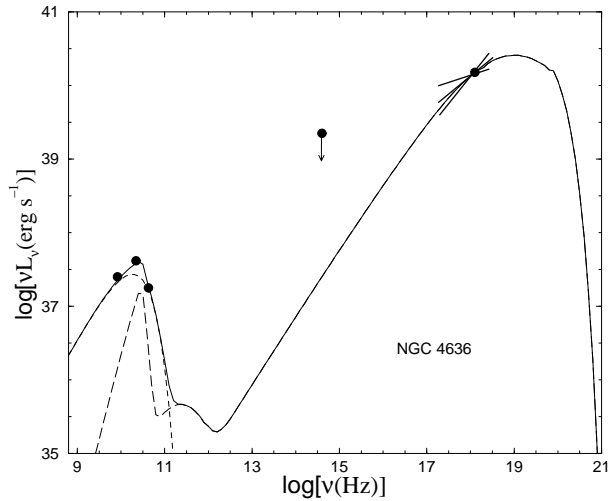,width=0.45\textwidth,angle=270}
\caption{The predicted spectrum of ADAF-jet model for 
NGC 4636 together with the observation. The long-dashed line is the 
spectrum of the truncated ADAF, the short-dashed line is for jet, and the 
solid line is the sum. The parameters are $\dot{M}
=3 \times 10^{-2} \dot{M}_{\rm Edd}, M=3 \times 10^8 \msun, Q_{\rm jet}=
10^{41.3} {\ergsec}, i=15^{\circ},
R_{tr}=100 R_g$, and $\gamma_e=300$.}
\end{figure}

We can imagine that there may exist such a case that the jet originates
at a moderate large radius so that the radio peak decreases a lot 
according to eq. (1) and it is the jet   
rather than the disk dominates the radio spectrum. This might be the case
for NGC 4636, as Figure 5 illustrates. 
For this source, the wind model predicts a synchrotron peak with 
a much higher frequency than expected by the radio and 
sub-mm measurements (Di Matteo et al. 2000).
Actually, in that model 
to fit the peak correctly, a black hole with much larger 
mass, $10^9 \msun$, than the observational value $3 \times 10^8 \msun$, 
must be used. Figure 5 shows this puzzle can be solved 
in the truncated ADAF model if
the truncation radius is as large as $100 R_{\rm g}$. 
To confirm this conclusion, polarization should be observed in radio band,
and the future low-frequency radio observation should 
not exhibit ``excess'' as in Sgr A$^*$, NGC 4649, and M87.  

\section{Summary and Discussion}
The predicted radio flux for nearby ellipticals by the canonical 
ADAF model is well
above the observations. If the mass accretion rate favored by the numerical
simulation were adopted, the ADAF model would also produce 
too much radio flux in 
the case of Sgr A$^*$. On the observational side,
strong evidence for the existence of jets in Sgr A$^*$ and 
these ellipticals are observed. Based on these facts, 
in this paper we propose an ADAF-jet model for Sgr A$^*$ 
and the nearby giant elliptical galaxies. 
We assume that  the advection-dominated 
accretion disk is truncated by a 
jet at a certain radius $R_{tr}$ within which 
most of the energy of the disk (if it still exists) is 
extracted for the powering of the jet therefore its radiation 
is completely suppressed. 
We calculate the spectrum of the ADAF-jet system for 
Sgr A$^*$ and three ellipticals and find that the radio overprediction 
are solved and the fits to the observation are
excellent. For example, for Sgr A$^*$, we can reproduce 
the break around the 50 GHz point in the radio wavelength, and 
the predicted size-frequency relation is in good agreement
with the observation. For M87, the truncation radius favored
by our model, $R_{\rm tr} \approx 24 R_{\rm g}$,
agrees with the most recent joint 
observation by VLBI, VLA and other instruments which shows
the jet should originate from within
$\sim 30 R_{\rm g}$ to the center black hole.

In our model we assume that the disk is radiatively truncated at 
radius $R_{tr}$ by the jet. It is interesting to note 
that this idea has been invented in the study to the black hole
X-ray binary GRS 1915 +105 where we have much better observational
constrains. Radio-infrared oscillation with 
periods of 10-60 min have been found to follow X-ray dips (Mirabel et al. 1998).
The oscillation has been found to be due to synchrotron emission
from repeated small ejections, i.e., jet, possibly resulted from suffering  
strong adiabatic expansion losses (Fender et al. 1997; 
Eikenberry et al. 1998), while the X-ray dips have 
been interpreted as the repeated {\em disappearance} of the inner accretion
disk (Belloni et al. 1997). 

The suppression of the radio flux also can be accessed by 
ADAF with strong winds (Blandford \& Begelman 
1999; Quataert \& Narayan 1999; Di Matteo et al. 2000). Our
truncated ADAF model is different with their wind model.
The wind model is based on the assumption that the Bernoulli parameter
of ADAF is positive therefore strong winds are inevitable. 
The validity of this assumption
is in debate at present. 
On the other hand, our model is based on the 
observation that jets really exist in the inner region of these sources
therefore may has more reliable physical basis. The predictions of these
two models are also different. Comparisons between the
prediction and observation on the location of the jet origin in M87, 
the $\approx 20\%$ variability at $\sim 1$ keV of M87, and the frequency-size
relationship of Sgr A$^*$, incline to support our model.

Noting that in nearby ellipticals and Sgr A$^*$, 
two kinds of sources where ADAF
is assumed to exist, jets are observed.
In NGC 4258, another source where ADAF is assumed to exist (Gammie, Narayan,
\& Blandford 1999), a jet also exists (e.g., Herrnstein et al. 1998).
Moreover,  in black hole X-ray binary a steady jet is usually 
present at the hard/low state, 
while this state is typically interpreted 
as an inner advection dominated accretion disk 
plus an outer standard thin disk (Fender 1999; 
Esin, McClintock, \& Narayan 1997). Thus, it seems that a jet 
always be present in an ADAF. If it is the case, we should  
check some result obtained under the frame of an canonical ADAF. 
An example is the radio/X-ray luminosity relation 
obtained by Yi \& Boughn (1998) in the frame of the standard ADAF. 
Since the radio spectrum below a possibly 
existed break frequency $\nu_{\rm break}$ (as in the sources 
presented in this paper)
may be dominated by the jet while that above a frequency the spectrum
may disappear due to the truncation of the disk at $r_{\rm tr}$, 
their conclusion holds only if the radio frequency 
$\nu$ satisfies,
\be
\nu_{\rm break} \la \nu \la 20 m_7^{-1/2}\dot{m}_{-3}^{1/2}T_{e9}^2 
\left( \frac{r_{\rm tr}}{20}\right)^{-5/4} {\rm GHz}.
\ee
Here $r_{\rm tr}=R_{\rm tr}/R_g$.

The crucial factor in our model is to assume that the
powering of the jet is the energy output from the disk 
rather than from the hole.
The excellent agreement between the predictions of our model
and the observations for both Sgr A$^*$ and 
elliptical galaxies in turn confirms the correctness of our assumption, and 
supports the ``disk'' origin of jets. 
In this context we note Livio, Ogilvie \& Pringle (1999) 
recently argue that since there is no reason 
to suppose that the magnetic field threading the central
spinning black hole differs significantly in strength from that 
threading the central regions of the disk, and the hole is a very 
poor conductor compared to the surrounding disk material, 
the electromagnetic output from the inner disk
region is expected to dominate over B-Z 
mechanism. We want to note that this conclusion doesn't contradict
the possibility that the rotation of the hole may play a role
in the powering of the jets.
This is because  a strong magnetic field may be created
due to the frame dragging in the ergo-sphere of the Kerr black hole, 
therefore the energy
extraction from the disk will be more efficient compared with 
the Schwarzschild black hole case (Koide et al. 2000).

\section*{Acknowledgments}
I thank Insu Yi and Junhui Zhao for the comments on this
paper and many helpful discussions with me. 
This work is partially supported
by China Postdoctoral Science Foundation.



\begin{thebibliography}{}
\def\refpar{\hangindent=3em\hangafter=1}
\def\reference{\bibitem{}}
\def\apj{ApJ}
\def\apjs{ApJS}
\def\mn{MNRAS}
\def\aa{A\&A}
\def\aas{A\&A Suppl. Ser.}
\def\aj{AJ}
\def\araa{ARA\&A}
\def\nat{Nature}
\def\pasj{PASJ}

\reference Abramowicz, M.A. 1998, in ``The Theory of Black Hole Accretion
Discs'', eds. M.A. Abramowicz, G. Bjornsson, 
and J.E. Pringle (Cambridge University Press)

\reference Abramowicz, M.A., Lasota, J.-P., \& Igumenshchev, I.V. 2000,
\mn, in press (astro-ph/0001479)

\reference Baganoff, F.K. et al. 2000, \apj, in preparation

\reference Belloni, T., Mendez, M., King, A.R., van der Klis, M., van 
Paradijs, J. 1997, \apj, 488, L109

\reference Blandford, R.D. \& Begelman M.C. 1999, \mn, 303, L1

\reference Blandford, R.D. \& Konigl, A. 1979, \mn, 232, 34

\reference Blandford, R.D. \& Payne 1982, \mn, 199, 883

\reference Blandford, R.D. \& Znajek, R.L. 1977, \mn, 179, 433

\reference Coker, R. \& Melia, F. 1997, \apj, 488, L149


\reference Di Matteo, T., Fabian, A.C., Rees, M.J., Carilli, C.L., \& 
Ivison, R.J. 1999, \mn, 305, 492 

\reference Di Matteo, Quataert, E., Allen, S.W., Narayan, R., \& Fabian, A.C.
2000, \mn, 311, 507

\reference Donea, A.C., Falcke, H, \& Biermann, P.L. 1999, in: 
"The Central Parsecs of the Galaxy", eds. H. Falcke, A. Cotera, 
W.J. Duschl, F. Melia, M.J. Rieke, ASP Conf. Series, Vol. 186, p. 162
(astro-ph/9909442)

\reference Duschl, W.J., \& Lesch, H. 1994, \aa, 286, 431

\reference Eggum, G. E.; Coroniti, F. V.;
\& Katz, J.I. 1988, \apj, 330, 142

\reference Eikenberry, S.S., Matthews, K., Morgan, E.H., Remillard, R.A.,
Nelson, R.W. 1998, \apj, 494, L61

\reference Esin, A.A., McClintock, J.E., \& Narayan, R. 1997, \apj, 489, 865

\reference Fabian, A.C. \& Rees M.J. 1995, \mn, 277, L55

\reference Falcke, H. 1999, 
in: "The Central Parsecs of the Galaxy", eds. H. Falcke, 
A. Cotera, W.J. Duschl, F. Melia, M.J. Rieke, ASP Conf. Series, Vol. 186, p. 113
(astro-ph/9909439)

\reference Falcke, H. \& Biermann, P.L. 1999, \aa, 342, 49

\reference Falcke, H., Mannheim, K., \& Biermann, P.L. 1993, \aa, 278, L1

\reference Fender, R.P. 1999, in ``Black Hole in 
binaries and galactic nuclei'' (astro-ph/9911176)

\reference Fender, R.P. Pooley, G.G., Brocksopp, C., Newell, S.J. 1997, \mn,
290, L65

\reference Ford, H.C. et al. 1994, \apj, 435, L27

\reference Gammie, C.F., Narayan, R., \& Blandford, R. 1999, \apj, 516, 177

\reference Ghisellini, G., Maraschi, L., \& Treves, A. 1985, \aa, 146, 204

\reference Herrnstein, J.R., et al. 1998, \apj, 497, L69

\reference Igumenshchev, I.V. \& Abramowicz, M.A. 1999, \mn, 303, 309

\reference Junor, W., Biretta, J.A., \& Livio, M. 1999, \nat, 401, 891

\reference Krichbaum, T.P., et al. 1998, \aa, 335, L106

\reference Koide, S., Meier, D.L., Shibata, K., \& Kudoh, T. 2000, \apj, in
press (astro-ph/9907434)


\reference Livio, M. 1999, Phys. Rep. 311, 225

\reference Livio, M., Ogilvie, G.I., \& Pringle, J.E. 1999, \apj, 512, 100

\reference Lo, K.L., Shen, Z.Q., Zhao, J.H., \& Ho, P.T.P. 1998, 
\apj, 508, L61

\reference Lovelace, R.V.E. 1976, \nat, 262, 649

\reference Lovelace, R.V.E., Ustyugova, G.V., \& Koldoba, A.V. 1999, 
in Proceedings of IAU Symp. 194, {\em Activity in 
Galaxies and Related Phenomena}.

\reference Mahadevan, R. 1997, \apj, 477, 585

\reference Mahadevan, R. 1998, \nat, 394, 651

\reference Manmoto, T., Mineshige, S., Kusunose, M. 1997, \apj, 489, 791

\reference Melia, F. 1992, \apj, 387, L25

\reference Melia, F. 1994, \apj, 426, 577

\reference Mezger, P.G., Duschl, W.J., \& Zylka, R. 1996, A\&AR, 7, 289

\reference Mirabel, I.F. et al. 1998, \aa, 330, L9

\reference Molteni, D., Lanzafame, G., \& Chakrabarti, S.K. 1994, \apj, 425, 161

\reference Nakamura, K.E. 1998, \pasj, 50, L11

\reference Narayan, R., Mahadevan, R., Grindly, J.E., Popham, R., \& Gammie, C.
1998, \apj, 492, 554

\reference Narayan, R., Yi, I., \& Mahadevan, R., 1995, \nat, 374, 623

\reference Quataert, E., \& Narayan, R. 1999, \apj, 520, 298

\reference Rawlings, S. \& Saunders, R. 1991, \nat, 349, 138

\reference Reynolds, C.S. et al. 1996, \mn, 283, L111

\reference Spruit, H.C. 1996, in `Physical Processes in Binary Stars', 
eds. R.A.M.J. Wijers, M.B. Davies and C.A. Tout, Kluwer Dordrecht, p249

\reference Stanger, V.J. \& Warwick, R.S. 1986, \mn, 220, 363

\reference Stone J.M., Pringle J.E., \& Begelman M.C. 1999, \mn,
310, 1002

\reference Yi, I. \& Boughn, S.P. 1998, \apj, 499, 198


\reference Yuan, F. 1999, \apj, 521, L55

\reference Yuan, F., Peng, Q., Lu, J.F., \& Wang, J.M. 2000, \apj, 537, 236

\end{thebibliography}
\end{document}